\documentclass[10pt]{article}

\usepackage{amsmath}
\usepackage{amssymb}

\usepackage{graphicx}

\usepackage{cite}

\usepackage{color}

\usepackage[colorlinks=true,urlcolor=blue]{hyperref}

\usepackage{geometry}

\usepackage{bm}

\usepackage[labelfont=bf,labelsep=period,justification=raggedright]{caption}

\makeatletter
\renewcommand{\@biblabel}[1]{\quad#1.}
\makeatother

\date{}

\begin{document}

\begin{flushleft}
{\Large
\textbf{Designing the statistically optimal drug for cancer therapy}
}
\\
\vspace{5mm}
Patrick N. Lawlor$^{1,2}$, 
Tomer Kalisky$^{3}$,
Stephen Quake$^{4}$,
Robert Rosner$^{5}$,
Marsha Rich Rosner$^{6,\ast}$,
Konrad P. Kording$^{1,2,\ast}$
\\
\vspace{2mm}
\footnotesize
1 Department of Physical Medicine and Rehabilitation, Rehabilitation Institute of Chicago, Chicago, IL, USA
\\
2 Department of Physiology, Northwestern University, Chicago, IL, USA
\\
3 Department of Engineering, Bar-Ilan University, Ramat Gan, Israel
\\
4 Department of Bioengineering, Stanford University, Palo Alto, CA, USA
\\
5 Department of Physics, University of Chicago, Chicago, IL, USA
\\
6 The Ben May Department for Cancer Research, University of Chicago, IL, USA 
\\
$\ast$ These authors made equal contributions
\\
E-mail: \href{mailto:m-rosner@uchicago.edu}{m-rosner@uchicago.edu}, \href{mailto:kk@northwestern.edu}{kk@northwestern.edu} 
\end{flushleft}

\section*{\normalsize \center{Abstract}}

\footnotesize

\leftskip=1.5cm \rightskip=1.5cm 

Cancer and healthy cells have distinct distributions of molecular properties and thus respond differently to
drugs. Cancer drugs ideally kill cancer cells while limiting harm to healthy cells. However, the inherent
variance among cells in both cancer and healthy cell populations increases the difficulty of selective drug
action. Here we propose a classification framework based on the idea that an ideal cancer drug should
maximally discriminate between cancer and healthy cells. We first explore how molecular markers can be
used to discriminate cancer cells from healthy cells on a single cell basis, and then how the effects of
drugs are statistically predicted by these molecular markers. We then combine these two ideas to show
how to optimally match drugs to tumor cells. We find that expression levels of a handful of genes suffice
to discriminate well between individual cells in cancer and healthy tissue. We also find that gene
expression predicts the efficacy of cancer drugs, suggesting that the cancer drugs act as classifiers using
gene profiles. In agreement with our first finding, a small number of genes predict drug efficacy well.
Finally, we formulate a framework that defines an optimal drug, and predicts drug cocktails that may
target cancer more accurately than the individual drugs alone. Conceptualizing cancer drugs as solving a
discrimination problem in the high-dimensional space of molecular markers promises to inform the
design of new cancer drugs and drug cocktails.

\leftskip=0cm \rightskip=0cm
\vspace{.5cm}

\section*{\large Introduction}
\normalsize

The central objective of treating cancer is to kill cancerous tissue while leaving healthy tissue intact.
Effective cancer drugs must therefore distinguish between cancer cells and healthy cells. Whether a cell
lives through the treatment depends on its ever-changing biological properties. Optimal cancer treatment
should also be robust to biological variability such as tumor cell heterogeneity \cite{cite1}. Combining all of these
ideas, we can frame the cancer problem in a way that balances the potential overlap of healthy and cancer
cell properties with the need to kill aggressive cancer cell variants. This framework should inform our
approach to treating cancer.

Cancer drugs should thus be conceived of as performing a computation on cells (Figure 1): Cellular
targets (e.g. particular cell surface proteins) lead to a single outcome (kill or do not kill) during treatment.
Mathematically, we can say that the effect of a drug is a mapping from a set of properties (targets of the
cell) onto a binary outcome (the cell lives or dies) - this is exactly the definition of a classifier in the fields
of statistics and machine learning \cite{cite2}. In that sense, any cancer drug is actually a classifier. Many
computer algorithms have been developed to solve classification problems and a rich literature exists in
the fields of statistics and machine learning regarding effective methods for classification \cite{cite2}. This
literature has come up with criteria for successful classification but these methods have not yet been used
to conceptualize the effect of cancer drugs (Figure 2).

\begin{figure}[!ht]
\begin{center}
\includegraphics[width=2.5in]{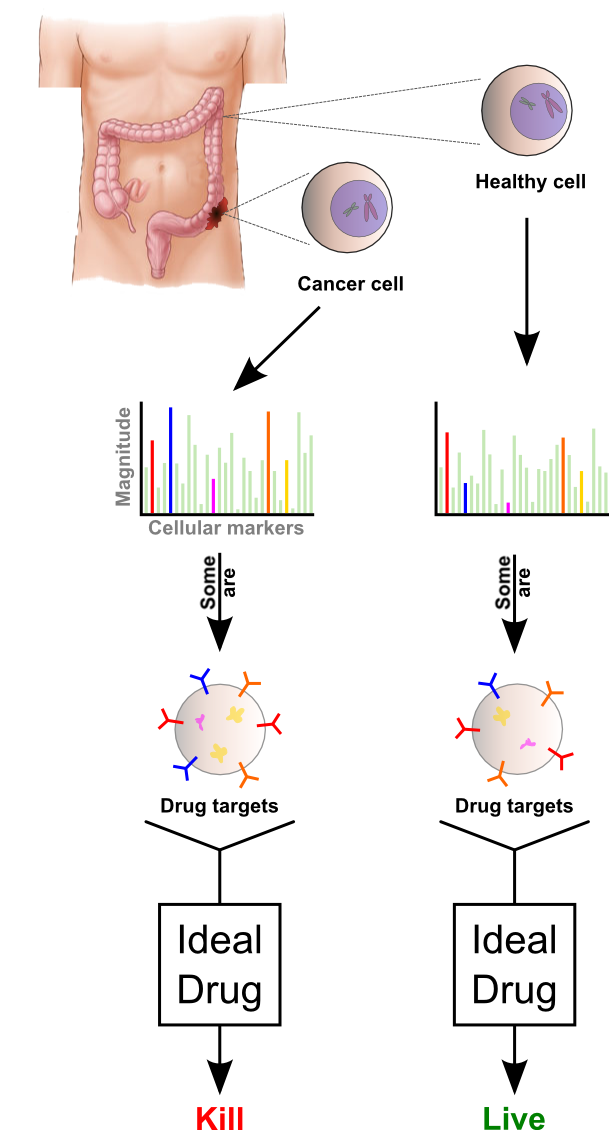}
\end{center}
\caption{
{\bf Cancer drugs solve a discrimination problem: kill (if cancerous) or no-kill (if healthy).}
}
\label{Figure_label}
\end{figure}

The properties of a cell are incredibly high-dimensional. On top of tens of thousands of genes with
millions of potential mutations, countless proteins in many configurations exist that can be located in
many areas of the cell. However, it seems possible that a small number of molecular targets will define
drug-cell interactions. In a number of tissues, a small number of aberrant cellular pathways contribute
most of the variance to a cell's identity as cancerous or healthy \cite{cite3}. Consequently, the expression levels of
a small number of genes may provide more than enough information to predict whether a cell is
undergoing unchecked growth or in a resting state.

Newly developed -omics approaches have accelerated our understanding of the complex relationship
between molecular biology and tissue phenotype. Microarray and sequencing technology, for example,
allow us to simultaneously collect information about thousands of cellular \emph{markers} – measurements about
the state of the cell, e.g., gene expression. Importantly, not all markers are drug targets. Molecular \emph{targets}
are molecules that cancer drugs actually use to alter cells. But perhaps buried in the thousands of 
measurable markers is a subset of markers that \emph{reflect or correlate with} the molecular targets of drugs.
For example, expression of genes that are downstream of a drug target may correlate well with that drug's
efficacy. Emerging biotechnology allows us to measure these cellular markers to more fully understand
cancer using statistical tools like machine learning.

Although cancer drugs have not been characterized as classifiers, machine learning has been extensively
applied to some aspects of cancer biology. One group has estimated breast cancer outcome by using
machine learning to create a 70 gene prediction algorithm \cite{cite4} while we and others have used machine
learning to predict metastasis-free survival \cite{cite5}. Others have attempted to distinguish between different
types of cancer using many types of algorithms including Support Vector Machines (SVM) \cite{cite6,cite7},
Principal Component Analysis \cite{cite8,cite9}, and Artificial Neural Networks \cite{cite10}. Yet others predict
chemosensitivity on the basis of gene expression \cite{cite11} and signaling networks \cite{cite12}. However, while all
these approaches have made impressive strides and are useful in clinical practice, the role of the cancer
drug itself has not been commonly conceptualized as solving a classification problem.

\begin{figure}[!ht]
\begin{center}
\includegraphics[width=4in]{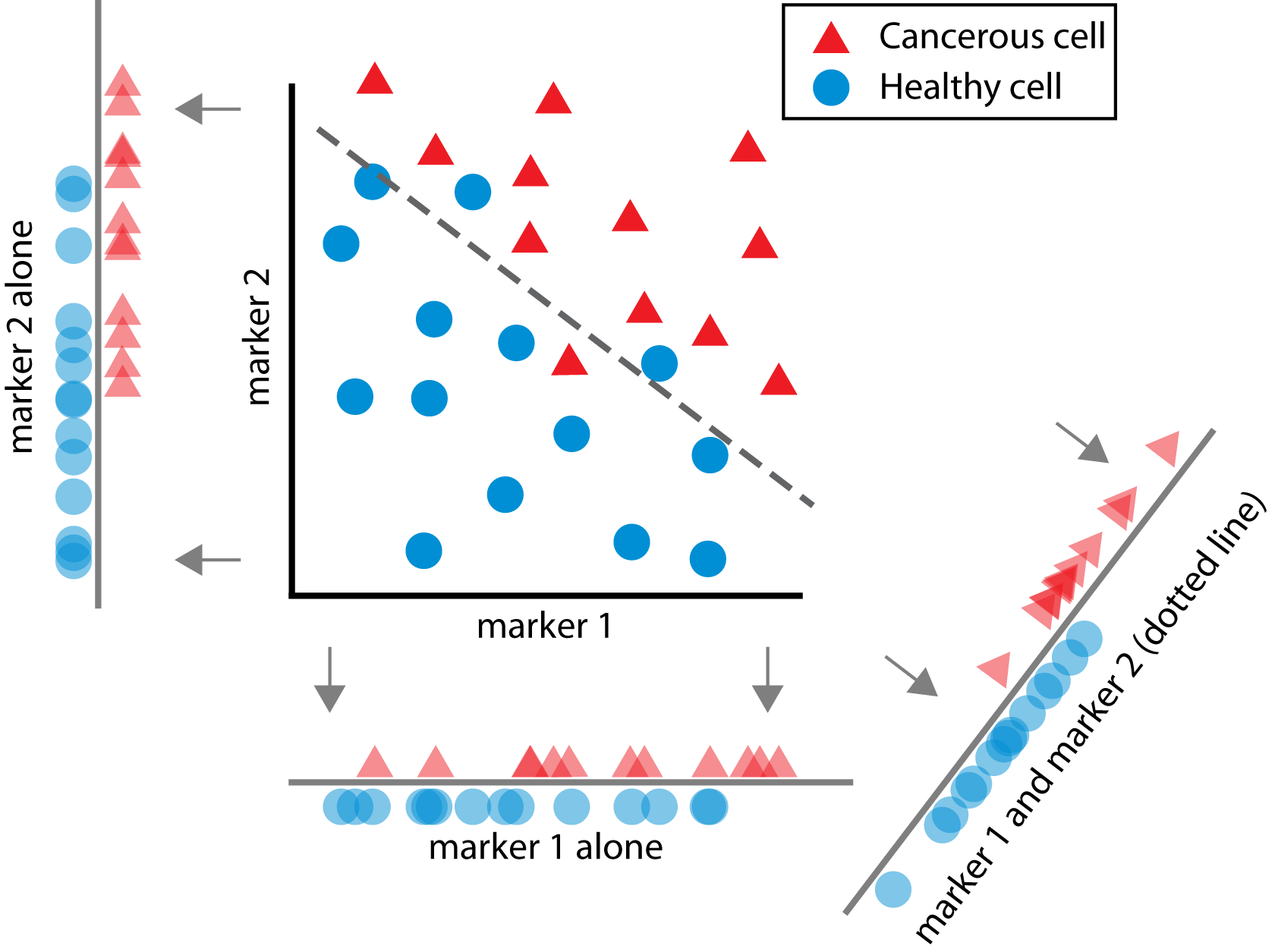}
\end{center}
\caption{
{\bf The idea of a classifier}, illustrating how one can combine information from two cellular
markers to construct a classifier that separates the two populations (cancerous and healthy cells) better
than either marker alone.
}
\label{Figure_label}
\end{figure}

The framing of cancer drugs as classifiers immediately begs a number of questions that we want to
answer here. (1) How classifiable are individual cancer cells versus healthy cells? Based on currently
measured cell markers, can the cancer versus healthy classification problem actually be solved? (2) How
much do actual drugs behave like classifiers based on properties of cells that are measured in current
experiments? (3) Can these statistical ideas be combined to enable drug discovery and optimization \cite{cite13,cite14}?

First, we use single-cell expression data to analyze the differences between cancerous and healthy tumors.
Next, we use cell line expression data to analyze how drugs actually classify cells to kill them or leave
them alone. We then combine these ideas to propose a novel approach to optimizing drug treatment.

\section*{\large Results}

\subsection*{\normalsize Part I: Markers for classification}

If cancer drugs are classifiers that use measurable markers as input, we can use standard classification
algorithms to ask if it is theoretically possible to solve the cancer versus healthy classification problem. If
it is theoretically possible, we would say that some optimal drug (or drug combination) would solve it.
This would imply that such a drug would kill all cancer cells while leaving all healthy cells alone. We will
use this notion of an optimal drug as a guide to analyzing treatment. In practice, actual drugs or drug
combinations should be chosen to resemble the optimal drug.

To ask if it is theoretically possible to solve this problem, we need a dataset of cells with both known
cancer state and measured markers. We thus used data from cancerous and healthy intestinal cells from
\cite{cite15} which contained the gene expression profile for each individual cell. For each of these cells we have
the expression of 45 genes: a 45 dimensional data vector. Can the classification problem from these
measurements to the cell state (cancer or not) be solved? To answer this question we used a standard
classification algorithm, Lasso GLM \cite{cite16}. Testing how well such classifying algorithms work allows us
to give an upper bound on how well an actual drug could work if it used gene expression alone.

\begin{figure}[!ht]
\begin{center}
\includegraphics[width=4in]{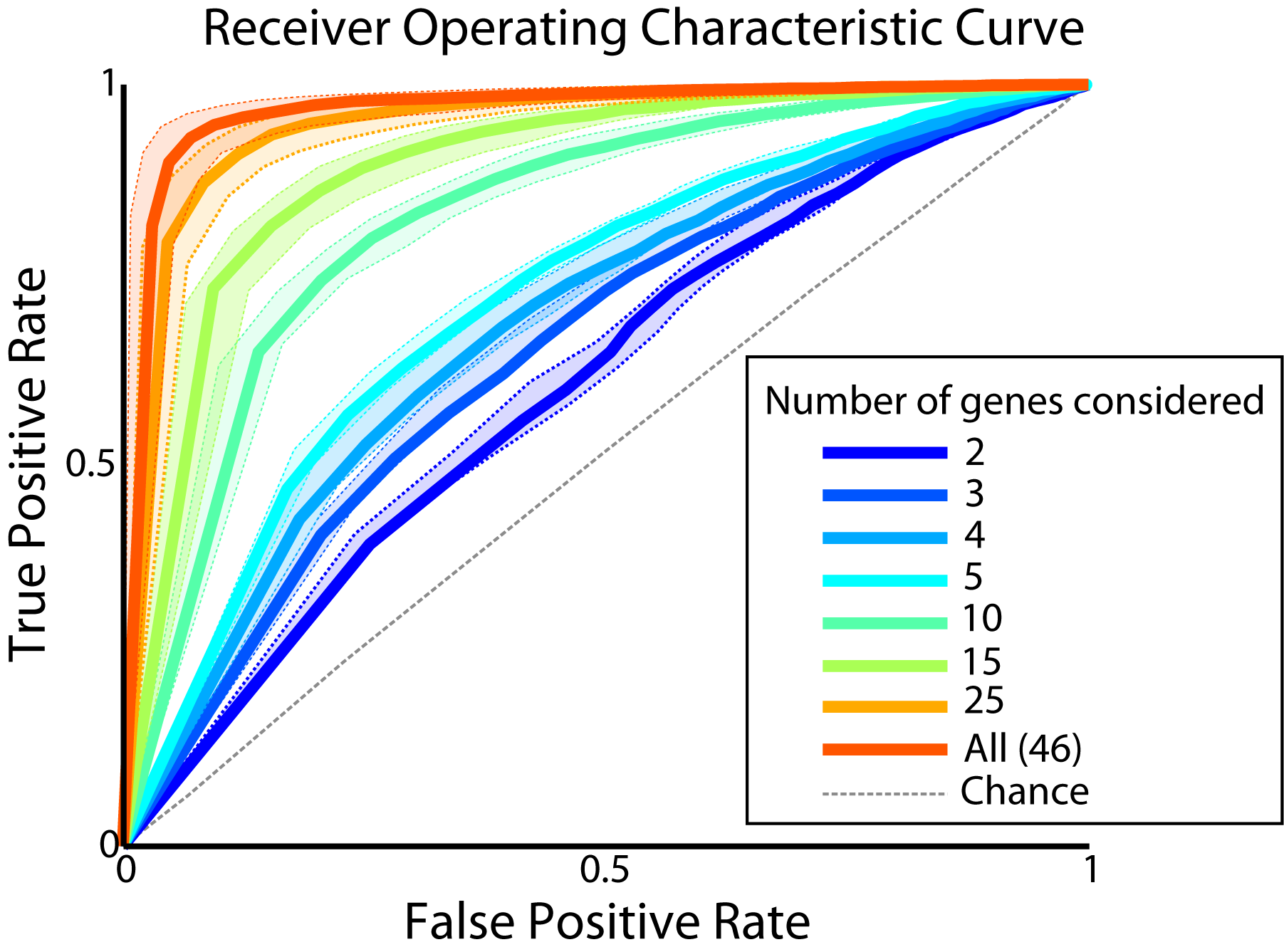}
\end{center}
\caption{
{\bf Performance of classifier as a function of the number of genes considered.} When measuring
accuracy of cell classification as cancerous or healthy, one should consider both types of errors: false
positives and false negatives (or more conventionally, true positives). This is illustrated by the Receiver
Operating Characteristic (ROC) Curve. Lines indicate median values, and shaded areas indicate 95\%
confidence intervals. Accuracy measured using cross-validation.
}
\label{Figure_label}
\end{figure}

First, we wanted to measure the possible accuracy of this classification. In classification there are
different kinds of errors that one can make. For example, it is easy to produce a drug that kills all cancer
cells but also kills all healthy cells. This drug would have 100\% true positives (killed cancer cells), but
also 100\% false positives (killed healthy cells). Thus, to fully characterize a classification strategy we
should analyze the relation of the two types of errors. The standard measure of classification accuracy is
the receiver operating characteristic (ROC) plot. In this plot the proportion of true positives is plotted as a
function of the proportion of false positives (Figure 3) to quantify both sensitivity and specificity. The
area under this curve (AUC) gives an overall measure of classification performance with a maximum
value of 1 for a perfect test. The AUC for our classification algorithm was $\sim$.9, indicating that healthy
and cancerous cells can be well classified. Therefore it is theoretically possible to solve the cancer versus
healthy classification problem for single cells with high accuracy using just a small set of 45 genes'
expression levels.

Next we asked \emph{how many} cellular markers an optimal drug would need to classify cells accurately. In
other words, what is the minimum number of markers an optimal drug must consider to tell the difference
between healthy and cancer cells? We found that a relatively small number of genes - approximately the
best\footnote{Here we define `best' according to the magnitude of the fit parameter corresponding to that gene. Gene expression data was standardized before analysis.} ten - sufficed to classify a cell as cancerous or healthy with high accuracy (Figures 3, 4). The
predictive power of the classifier saturated soon after the best ten genes were included. Thus, only a small number of cellular markers provide the majority of the information used to classify a cell as cancerous or healthy.

In this section we have analyzed data from single cells; how would our results have been with cell lines
instead? Expression data from cell lines reflects the group averages which can decrease noise, but can
also obscure the effect of outliers and other variability. To examine the effects of this heterogeneity, we
applied the above analysis to expression measurements made on cell lines classified as more or less
invasive (luminal or non-luminal)\footnote{Data set used here is described in more depth in section II and in the Methods section.}. We found that we could classify single cells as cancerous with an
AUC of $\sim$.9 (see above) whereas we could perfectly classify cell lines as more or less invasive (AUC of
1). This suggests that cell-line expression data is easier to classify as it lacks the heterogeneity observed in
single cells. Thus, we should approach treatment on the basis of single cell data rather than population
data to ensure that we are targeting all the relevant cells.

The ability to measure gene expression from single cells raises the question of whether it is more
important to measure more cells or more genes. To answer this, we quantified the relative importance of
increasing the number of measured cells versus the number of measured genes per cell. We trained the
classifier with numbers of cells ranging from two total (one healthy, one tumor cell) to 188 total cells (all
cells). As above, we measured the classifier's performance, except that we did so for each training
scenario. We again found that the performance saturated after a small number of genes for each training
scenario. Importantly, we also found that performance continued to improve with increasing numbers of
training cells until approximately 80 were used. Thus, measurements from at least tens of cells are
required to characterize a population of tumor and healthy cells.

\begin{figure}[!ht]
\begin{center}
\includegraphics[width=4in]{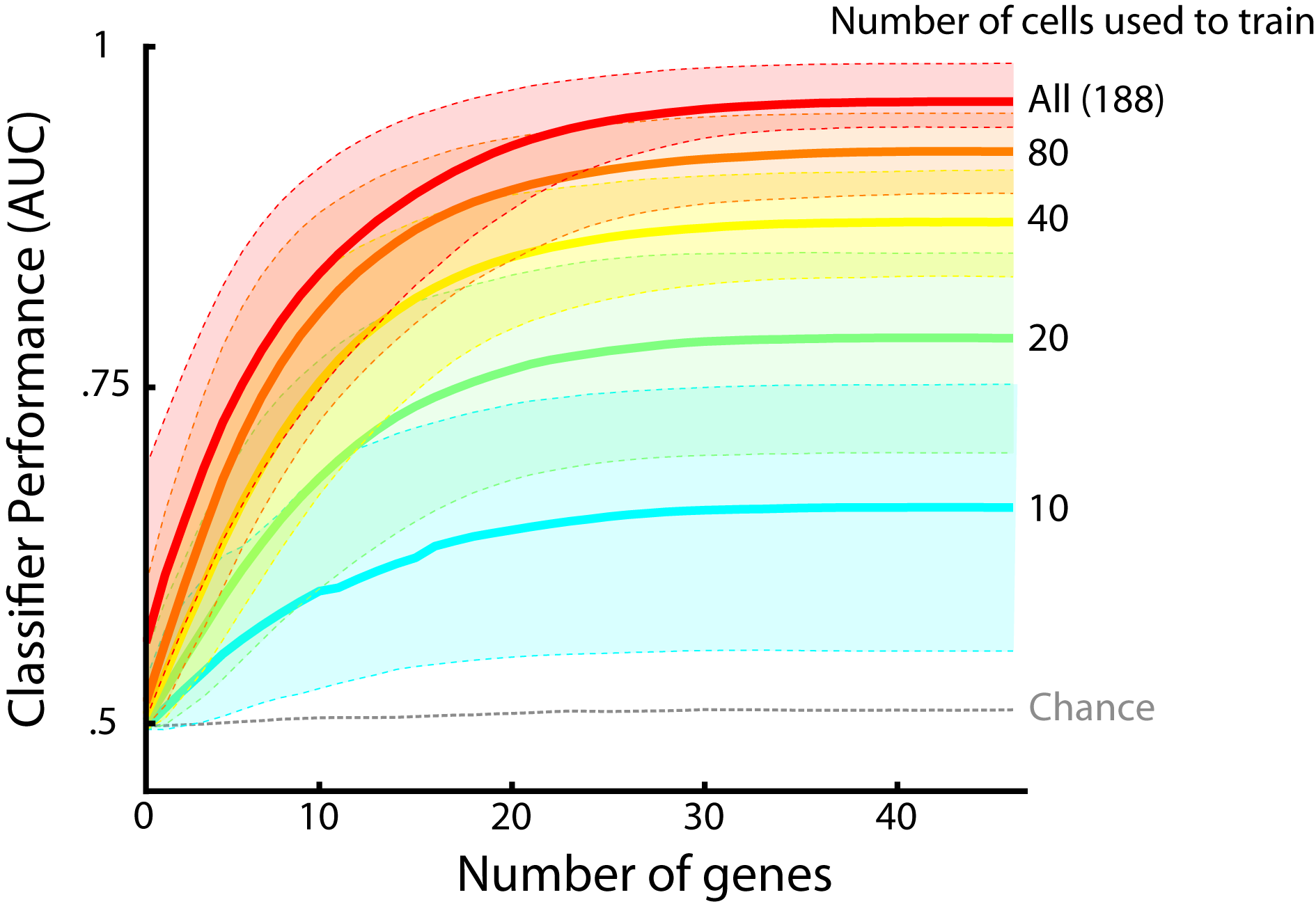}
\end{center}
\caption{
{\bf Performance of the classifier as a function of the number of cells and genes measured.}
Classifier performance was measured as area under the curve (AUC) of the ROC curve. Each colored line
represents a different number of cells used to train the classifier. Lines indicate median values, and shaded
areas indicate 95\% confidence intervals. Accuracy measured using cross-validation.
}
\label{Figure_label}
\end{figure}

\subsection*{\normalsize Part II: Classification by real cancer drugs}

In the first part of this study we have explored how an optimal cancer drug could tell the difference
between healthy and cancer cells. In this second part of the study, we quantify the dependency of drug
response on markers by pooling all cell lines (more and less invasive). We then ask how cancer drugs
actually relate to markers. To allow us to ask this question, we used a set of cell lines with gene
expression data as well as cancer drug sensitivity measurements from \cite{cite17}. We use an algorithm to
predict if cancer drugs will kill cells of a specific cell line, given its markers.

An optimal drug could use cellular markers to classify cells, but do actual drugs behave this way? If so,
these markers should predict the probability that a drug kills the cell: its chemosensitivity. Unlike the
optimal drug, however, actual drugs will not necessarily target cancer cells in an ideal way. An actual
drug will target cells based upon its mechanism, which may not exactly align with the ability to
distinguish cancerous from healthy cells. Such a suboptimal drug could still be a classifier, however, if it
uses molecular markers as inputs to make a binary decision, even if that decision is the wrong one.
Alternatively, a drug may make the decision to kill a cell in a way that is not predictable by measurable
markers, and therefore wouldn't be an apparent classifier. For example, a drug could kill cells in a way
that depends only on its concentration and not on the state of the cell. Thus, we will first ask if actual
drugs behave as classifiers at all.

We tested how well gene expression does predict drug response. Because there currently exist no data
linking single cell markers to drug responses, we instead made predictions about cell lines. In particular,
we estimated the drug concentrations needed to kill cells in these lines. Indeed, aspects of the drug
responses were predictable. For example, we predicted the chemotherapeutic response to the drug
Lapatinib, with an R value of $\sim$0.6. This roughly means that gene expression predicts 60\% of the
variability in chemosensitivity. However, the low R value shows that our predictions of drug behavior
were not strong. This may imply that not all relevant markers were measured, or alternatively that the
relationship is nonlinear and not captured by linear machine learning methods. While molecular markers
such as gene expression do not capture all variability, they do appear to play a significant role in
predicting the drug response. Therefore, cellular markers predict some degree of drug behavior when
treated as inputs to a classifier.

\begin{figure}[!ht]
\begin{center}
\includegraphics[width=3in]{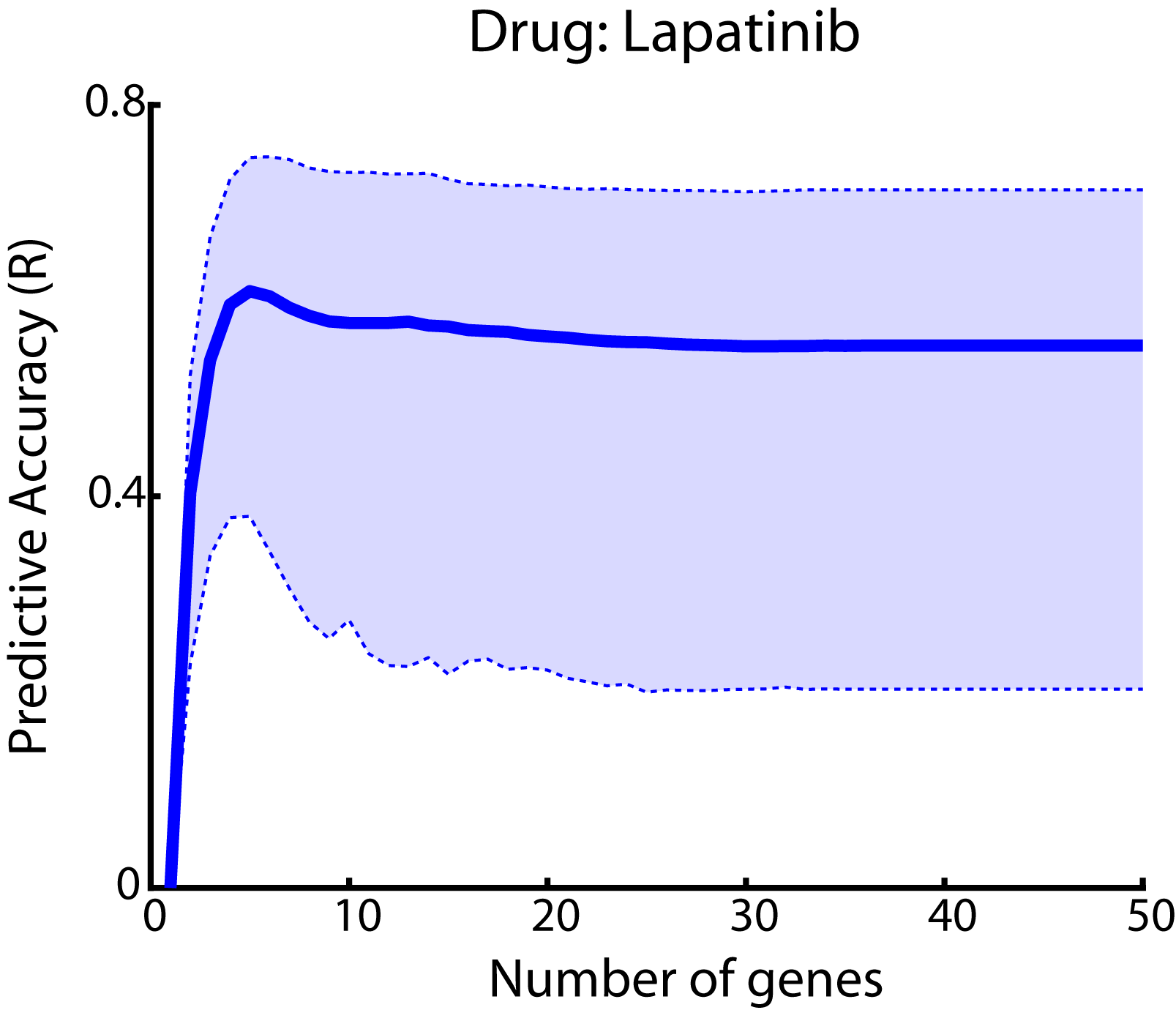}
\end{center}
\caption{
{\bf Cancer drug classification as a function of the number of genes considered.} Accuracy of
chemosensitivity prediction, measured as the coefficient of determination, R. Accuracy measured using
cross-validation.
}
\label{Figure_label}
\end{figure}

Although we found that a small number of genes could theoretically predict whether a cell was cancerous,
we wanted to know how many genes suffice to predict actual drug behavior. To do this, we measured the
drug's behavior as a function of the number of genes. Similar to an optimal drug, Lapatinib needed only
$\sim$15 genes to reach its peak accuracy (Figure 5). Thus, a small number of cellular markers predict whether
or not actual drugs kill a cell.

If drugs act as classifiers using a small number of properties, then a drug can be characterized by plotting
its effect as a function of the properties. We thus selected the two genes (PERLD1 and SF3B3) that were
jointly best at predicting Lapatinib's effect across cell lines. Using simple grid interpolation and
extrapolation routines we plotted the drug's effect as a heatmap (Figure 6). This type of visualization
shows that even two genes can capture the complex behavior of a cancer drug.

\begin{figure}[!ht]
\begin{center}
\includegraphics[width=4in]{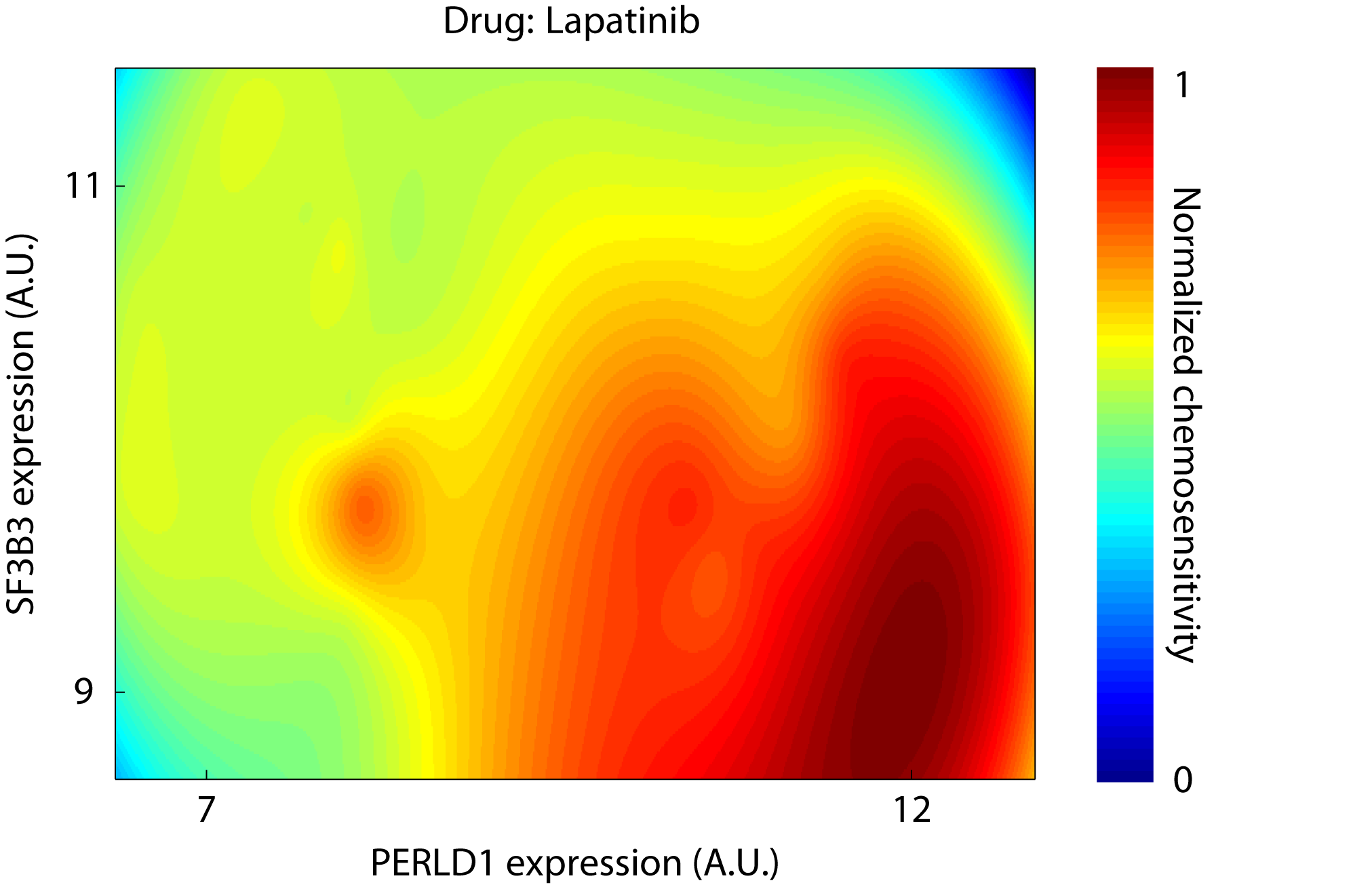}
\end{center}
\caption{
{\bf Drug sensitivity heatmap as a function of the two most important genes.}
}
\label{Figure_label}
\end{figure}

\subsection*{\normalsize Part III: Optimizing cancer treatment}

In the previous two sections we have shown how an optimal cancer drug could classify cells as cancerous
or not, and that cellular markers predict how actual drugs behave. In this section, we combine these two
ideas to outline a possible approach to optimizing cancer treatment. Arguably, there cannot exist cell lines
of healthy cells, so we treat invasiveness (luminal versus non-luminal) as a stand-in for this statistical
problem. We thus use the same data set as in section II, which includes gene expression data as well as
drug sensitivity data for a group of cell lines. We start with one given suboptimal drug that does not
distinguish well between invasive and non-invasive cells. We then asked, which marker promises to be
the best target to be added. In other words, which combination of drug and a single extra marker, in
theory, can best distinguish invasive from non-invasive cells?

Although actual drugs are often far from ideal, we can use machine learning to describe how to make any
given actual drug more like the optimal drug. This is inspired by the well-known approach in machine
learning called boosting \cite{cite18}, in which additional features are added to a classifier to enable progressively
better performance. More specifically, given (1) the optimal classification of cells (in terms of gene
expression) and (2) the actual drug's behavior (also in terms of gene expression), we can determine which
gene would improve the actual drug the most if it were targeted. To do this, we again use the GLM
framework to say that the optimal drug's effect depends on a combination of the actual drug and a given
gene. We then iterate through all possible genes to determine which single gene best approximates the
optimal drug when combined with the actual drug. This idea allows us to determine the next best
molecular target.

As an example to demonstrate how to computationally determine the next best molecular target, we used
the data set from section II. We defined the optimal strategy as preferentially targeting non-luminal
(invasive) cancer over luminal (less invasive) cancer, and first asked how the optimal drug would
distinguish these two classes of cancer. We then chose Lapatinib as the actual drug. Next, we found a
single target gene that would improve Lapatinib's classification (Figure 7). Importantly, the drug and
marker together provide a better classification than either alone (compare with Figure 2). Thus, Lapatinib
could be improved by additionally targeting the marker, TMEM106C. To produce optimal drug cocktails
we would try combinations of Lapatinib with drugs known to relate to TMEM106C.

\begin{figure}[!ht]
\begin{center}
\includegraphics[width=6in]{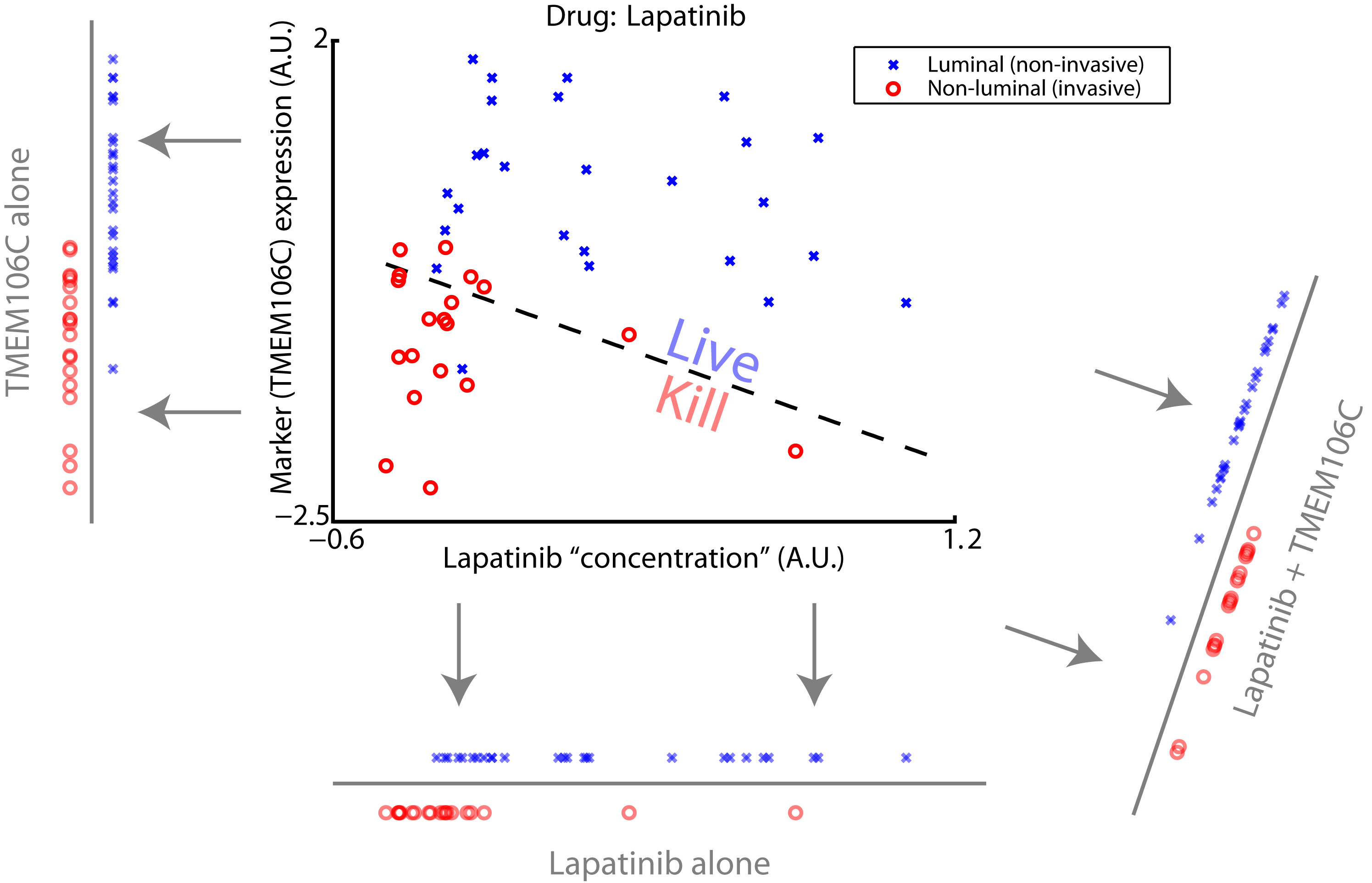}
\end{center}
\caption{
{\bf Cancer treatment optimization.}
Better discrimination between cell populations is achieved by
including an additional cellular marker with the drug, Lapatinib. See text for full description.
}
\label{Figure_label}
\end{figure}

We find that targeting one additional marker significantly improves the accuracy of classification. A
cocktail consisting of Lapatinib and a drug whose action relates to TMEM106C improves classification
accuracy nearly to the level of the ideal drug (Figure 8, second bar). Targeting other markers – the next
best markers – can also improve performance considerably above Lapatinib alone (Figure 8, bars 3 -10).
Thus our method provides a method to choose additional drug targets in a way that allows us to target
cancer cells more accurately.

\begin{figure}[!ht]
\begin{center}
\includegraphics[width=4in]{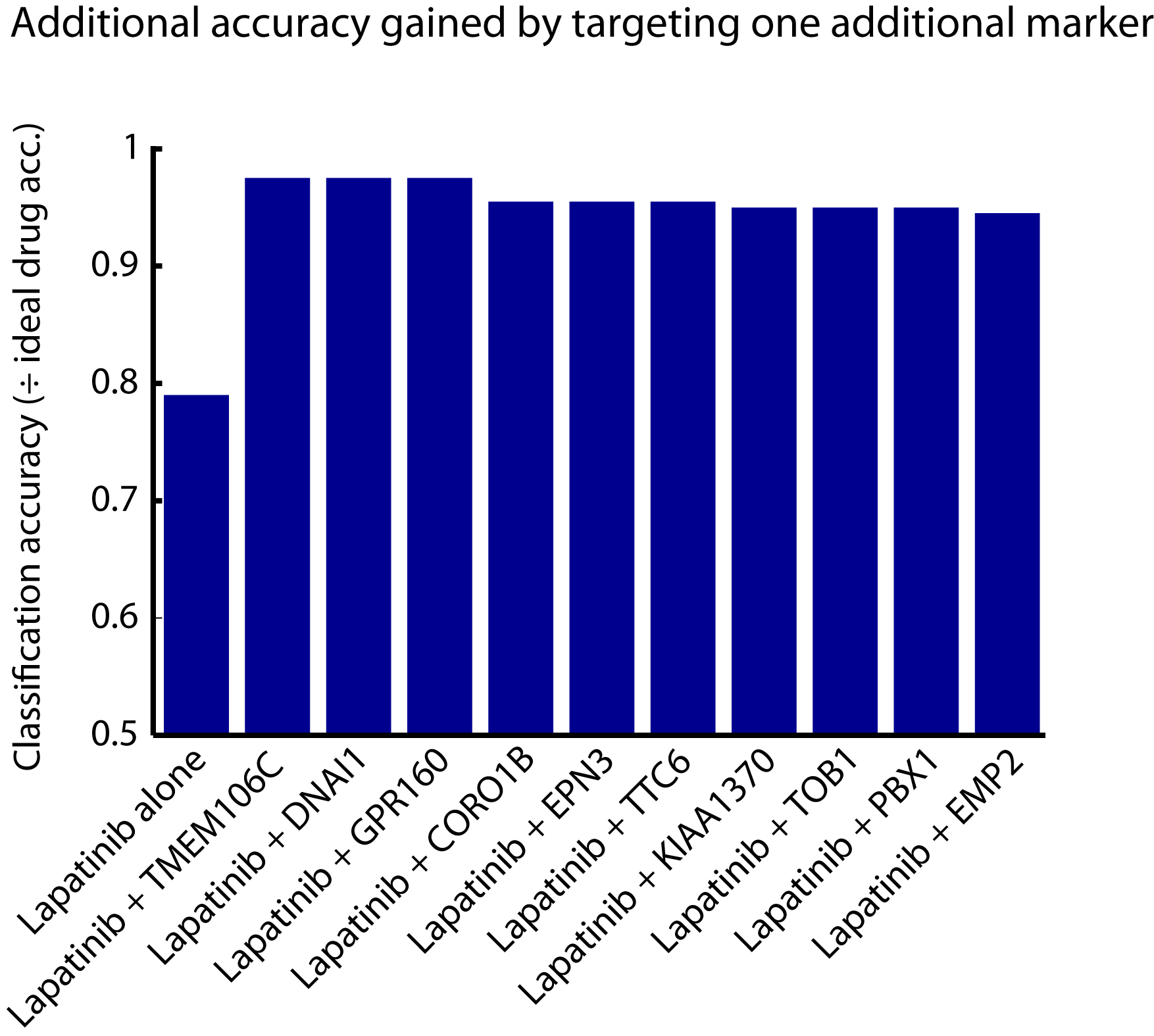}
\end{center}
\caption{
{\bf Classification accuracy showing the improvements achieved by targeting one more marker.}
}
\label{Figure_label}
\end{figure}

\section*{\large Discussion}

In this study, we argued that cancer drugs can be conceptualized as classifiers. Optimal drugs use
molecular targets to kill cancer cells while minimizing harm to healthy cells. We framed this problem as
one that could be addressed with tools from machine learning and explored its implications. We found
that gene expression alone, one class of molecular markers, could indeed be used to solve this
optimization problem quite well. We also found that actual drugs behave like suboptimal classifiers.
Finally, we suggested ways of using the classification framework to derive drug development strategies
that perform as closely as possible to an optimal drug. Thinking of cancer drugs as classifiers provides a
conceptual framework for devising optimal treating strategies for cancer.

The distinction between molecular \emph{markers} (i.e., phenotypic characteristics of a cell) and molecular
\emph{targets} (i.e., cell components such as proteins whose functions are altered by a drug) is crucial. Gene
signatures are composed of molecular markers that in some cases may be drug targets. We argue here
that drugs can act as classifiers by using gene signatures, but note that the corresponding gene transcripts
and proteins coded by these genes generally are not the direct targets of the classifying drug. Instead, the
genes predictive of a drug's action may be indirectly related to the drug's actual targets by way of a
shared biological network, for example. This study strictly observes the distinction between cellular
markers and targets.

One limitation of this study is that gene expression alone is unlikely to entirely predict something as
complicated as drug response. Cell behavior and phenotype are the results of complex chemical and
physical interactions between genes, epigenetic factors, proteins, and the local environment. Current
technology does not allow measurement of many types of markers of interest, and this amount of data
would be overwhelming even if it were possible. In light of how little information gene expression
represents, it is perhaps surprising that it predicted drug behavior reasonably well.

Additionally, the two datasets we used measured markers of different types of cells. Unlike the Quake
dataset (section I), the Gray dataset (sections II and III) measured expression of cultured cell lines rather
than expression levels in primary single cells. Therefore, those gene expression levels reflect a population
average of the particular cell line rather than the expression of individual cells. It is possible and of great
therapeutic importance that small numbers of drug-resistant cells with distinct gene profiles may not be
detectable when the whole tumor cell population is analyzed together. For example, small populations of
cells within a tumor have been shown to drive the evolution and drug resistance of some types of cancer
\cite{cite1,cite19,cite20}. Thus, an important question that emerges from this study is whether experiments with tumor
cell \emph{populations} capture as much relevant information as experiments with \emph{individual cells}. Here we use
the Gray dataset of cell lines as stand-ins for single cells, but acknowledge that single-cell treatment data
would be optimal. The Gray dataset was the best available dataset for our chemosensitivity study, but
more rigorous single-cell treatment analysis will have to wait for that experimental data to become
available.

Another difference between the two datasets we used is that they come from different organs. The first
section uses a dataset from intestinal tissues whereas the second and third sections use a dataset from
breast cell lines. This no doubt makes it difficult to directly compare molecular markers across sections of
this study's analysis. However, some properties of cancer are likely to be true regardless of particular
organ system \cite{cite3}. If cancer drugs behave as classifiers for colon cancers, they will likely behave as
classifiers in other organs. Analyzing cells originating from different tissues is therefore not likely to
affect the conclusions of this study.

In this study we propose a method to optimally choose potential drug markers for combination drug
regimens. In practice, after choosing an initial drug, this would consist of computationally finding the
next best potential drug marker and then targeting this marker. One possible way to clinically target the
next best marker would use the results of this study. Section II tells us which genes predict the behavior
of a given drug. If we know which gene we would like to target, we will simply begin by looking for
drugs whose behavior is predicted by that gene. Empirical testing of multiple drugs could then determine
which single drug is best and in what dose.

This approach makes sense if the combined drugs don't affect one another. For example, it may be that
the drug targeting the `optimal' additional marker TMEM106C significantly interacts with the molecular
mechanism of Lapatinib, and vice versa. If the drugs do not interact, i.e., the effects of the individual
drugs are additive, the ability of Lapatinib to classify cancer cells would not be affected by the additional
drug. Thus the compound classifier – the drug combination – would classify cancer cells more accurately
than either drug alone. It is also possible that weak nonlinear interactions between drugs could still yield a
superior compound classifier than either drug alone, but its performance would likely be suboptimal.
Assuming additivity places a \emph{theoretical upper bound} on how well drug combinations could work.

What if we cannot find a drug that both targets TMEM106C and does not interact with Lapatinib? This
will likely occur in practice. One solution would be to iteratively target the next-most-optimal markers
(Figure 8) until we find a drug that targets a semi-optimal marker that does not interact with Lapatinib.
For example, marker $A$ (and its corresponding drug $A_D$) might classify cancer cells more accurately than
marker $B$ (and its corresponding drug $B_D$), but if drug $A_D$ interacts undesirably with Lapatinib, but drug
$B_D$ does not, then marker B would prove to be the better choice of target. Thus, choosing the next best
marker to target may require theoretical and empirical considerations. This method cannot guarantee that
a given combination of drugs will work but instead proposes a more efficient way to select drug cocktails
for testing.

Another way to address this problem is to consider sequential as opposed to simultaneous drug treatment.
Initial cell profiles can be used to identify drug $A$ as the optimal single drug. Once treated with drug $A$,
the gene expression profiles of these cells will change. If one then utilizes the altered gene expression
profiles in cells that have been treated with drug $A$ to identify the next optimal drug, $B$, then the two drug
treatments are by definition independent. Interestingly, a recent study by Yaffe and colleagues showed
that the killing of aggressive, triple-negative breast cancer cells was more effective when two drugs were
applied sequentially as opposed to delivered in combination \cite{cite21}. By approximating the optimal treatment
strategy using sequential drug administration one may avoid the complications of drug interactions.

Many previous cancer studies used statistical approaches in conjunction with large datasets. The Friend
group, for example, used machine learning to predict prognosis and phenotypes in breast cancer \cite{cite4}. This
resembles our strategy in that they use gene expression to make predictions about cancer. Other groups
have used machine learning to distinguish between sub-types of cancer \cite{cite6}. They typically combine
microarray data with such algorithms \cite{cite7,cite22} to distinguish between types of cancer. This resembles our
strategy to optimally discriminate between cell types. The Golub group used machine learning to classify
cell lines as belonging to a drug-resistant or drug-sensitive class based upon gene expression \cite{cite11}. This
resembles our strategy of modeling drug responses. Finally, the Nowak group outlined a way to design
drug cocktails that are robust to genetic mutation \cite{cite23}. We extend these lines of research, highlighting a
process by which potential targets for drug optimization can be identified by \emph{defining discriminability as
the drug's objective}. Furthermore, we demonstrate the importance of using single cell data, as opposed to
pooled cell population data, to solve the classification problem.

Conceptualizing drugs as classifiers is not only meaningful for cancer treatments. Any drug that should
produce a binary outcome could be modeled using the same framework. This framework should
generalize to drugs that are supposed to attack pathogens, such as malaria, or undesirable cells, such as
those constricting breathing during asthma attacks. Such applications may actually be easier than the
applications possible in cancer biology.

Thinking of drugs as classifiers paves the way for advances in the treatment of cancer. Since cancer cells
are discriminable from healthy cells on the basis of gene expression or other molecular markers, we could
use molecular biology to design extremely specific drugs. In essence, one would design a drug that would
require cells to meet multiple conditions before being destroyed \cite{cite24}. For example, these conditions could
simply correspond to the expression levels of the ten genes most predictive of cancer. Many cells would
likely meet one or two of the conditions but exceedingly few would meet all of them, ensuring that the
drug only destroyed the cells it was designed to destroy. Although no such drug has yet been created,
designing drugs to perform classification will yield treatments that far surpass the abilities of current
drugs and with fewer side effects.

Importantly, this framework allows the possibility of optimally designing cocktails of cancer drugs under
the assumption of weakly interacting drugs. Although each cancer drug may preferentially attack cells
with certain markers, it is unlikely that each human cancer will respond well to a single drug. But given a
particular cancer's measurable markers and knowledge of drugs' preferred markers, it should be possible
to estimate a cocktail of drugs whose net preferred marker profile closely matches the cancer's marker
profile. Here, we have described one way to approach this, and other methods of optimization are likely
possible. Such a strategy could actually be implemented rather quickly as it would only require drugs that
are already approved for treatment.

Current approaches focus on exploring the efficacy of individual drugs and drug combinations, but
without an explicit framework for defining and constructing optimal drugs. Our approach allows one to
define optimal drug performance, and outlines a strategy by which drug combinations can be devised that
may optimally approach the (theoretical) optimal drug performance.

\section*{\large Materials and Methods}

In this paper, we frame cancer drugs as solving a classification problem. As such, we approach this
problem with the tools of machine learning, a field that routinely addresses these types of questions. Here
we apply the Generalized Linear Model (GLM) framework, a machine learning algorithm, to explore the
behavior of cancer using two gene expression data sets.

\subsection*{\normalsize Data}

We used one dataset to explore how cancerous and normal tissue differ \cite{cite15}. It was collected by Dalerba
et al and provided by Tomer Kalinsky and the Quake lab. They measured gene expression by performing
PCR on single cells isolated from healthy intestine, primary intestinal tumor, and a xenograft derived
from primary intestinal tumor tissue. Because their apparatus allowed them to measure the expression of a
relatively small number of genes ($\sim$50), they chose these genes related to a property they were interested
in: stem cell-like properties. More specifically, they chose them on the basis of a) reported association
with stem cell-like properties in the literature and b) an iterative method that identified genes that were
co-expressed with identified markers of stem cells. This means, that the genes were optimized to
distinguish stem-cells from non-stem cells and not cancer from non-cancer cells, making our machine
learning problem more difficult.

Because differences between healthy and cancerous tissues may be subtle, we sought to explore
differences between the most similar groups of cells we could find in the data set. The original study used
principal component analysis to identify sub-populations within the healthy, primary tumor, and
xenograft tissues. We chose one particular sub-population that existed in both healthy and primary tumor
tissues: stem-like cells. We then tested whether we could use a GLM to reliably identify which cells
belonged to the healthy tissue and which belonged to tumor tissue, and how many properties were needed
to do so.

We used another dataset to ask if actual cancer drugs act as classifiers and to determine how to optimize
cancer treatment. This was collected by Gray, et al. for \cite{cite17}. They measured the expression of
approximately 19,000 genes in breast cancer tissues using microarrays, and the chemotherapeutic
responses of those tissues. The breast cancer tissue came from a panel of approximately 50 breast cancer
lines. After gene expression of each cell line was measured, each line was treated with one of 77 drugs.
They defined the sensitivity of a cell line to a given drug as the concentration of drug at which 50\% of
cell growth was inhibited (more specifically, they took the negative log of this number). We thus have a
dataset where we know the markers (before treatment) and we know how strongly the cells responded to a
variety of drugs.

Within the Gray dataset, we also needed to define ``cancerous'' and ``non-cancerous'' cell lines despite the
fact that nearly all cell lines were actually cancerous. We thus divided the 50 cell lines into malignant and
benign cell lines based upon whether they were luminal in origin (benign) or not (malignant). This had
already been done in \cite{cite17}. Grouping the cell lines in this way allowed us to formalize targeting one
distinct population over another within that dataset.

\subsection*{\normalsize Analysis}

To extract useful information from these large datasets we used the Lasso Generalized Linear Model
(Lasso GLM) framework implemented in Matlab. GLM's are a class of machine learning algorithms that
extend least-squares regression to target variable distributions other than the normal distribution. GLM's
essentially relate a linear combination of predictor variables (like gene expression) to the predicted
variable (like probability of being a cancerous cell) by passing the linear combination through a special
function, the inverse link function, $g^{-1}$. This link function is chosen to reflect the distribution of the
predicted variable. The variable of interest, $f$, is related to the predictor variables, $\mathbf{x}$ (e.g., a vector of
genes' expression levels), and their weights, $\bm{\beta}$ (or $\beta_k$ for each $k^{th}$ gene), by:
\\
\[
f(\bm{\beta},\mathbf{x}) = g^{-1}\left(\beta_0 + \sum_{k}x_k\beta_k\right)
\]
\\
To prevent over-fitting, we implemented L1 regularization. Regularization penalizes the algorithm for
choosing a model that is too complicated – i.e., one with too many predictive features (genes). This forces
it to choose only genes that contribute to goodness of fit. Regularization helped avoid fitting noise rather
than signal, thus allowing us to capture the essential information in the data.
During estimation of the model parameters, L1 penalizes choice of $\bm{\beta}$ in proportion to its absolute value:
\\
$$
\text{penalty} \sim \sum_k \left| \beta_k \right|
$$
\\
\subsection*{\normalsize Classification}

In the first part of the study, we build a GLM to classify cells as cancer or healthy using single-cell PCR
data. We use a GLM with the Bernoulli distribution. That is, we say that a cell either came from the
healthy population (a value of 0) or from the cancerous population (a value of 1) but nowhere in between.
Given the gene expression profiles for cells known to be healthy or cancerous, the GLM returns a weight,
$\beta_k$ for each $k^{th}$ gene according to how predictive that gene is.

$$
\text{cancer state}(\bm{\beta},\mathbf{x}) = \frac{1}{1 + exp(-\mathbf{x}\bm{\beta})} 
$$

To ensure that our classifier generalized to data beyond training data, we used cross-validation (CV). We
performed two levels of CV: 10-fold CV to determine the optimal value of the regularization penalty
parameter (within the training set only), and 10-fold CV to test the predictions of the model. In this
approach, we used 90\% of the data to train the algorithm, and compared predictions made by that 90\% to
the actual values of the remaining naive 10\%. We performed this ten times total, using a different 10\% of
the data for each round so that all of the data was eventually used as both training and validation data.
This allowed us to show that our classifier worked with arbitrary data rather than just data it had already
seen.

To quantify the accuracy of the GLM's predictions, we constructed Receiver Operating Characteristic
(ROC) curves. An ROC curve characterizes the sensitivity and specificity of the classifier. Another way
to quantify the overall performance of the prediction is to measure the area under the ROC curve. A
perfect predictor algorithm would achieve an ROC integral value of 1, whereas a random predictor would
achieve an ROC integral value of 0.5. ROC curves provided a straightforward way to interpret the
accuracy of our classifier.

To measure the relative importance of measuring more markers versus more cells, we used smaller sets of
data to train the classifier. We did this by randomly choosing training data points for each of 100
iterations, for each training set size. For example, for a training set size of four points, we chose a
different set of four training points each of 100 times. Then for each set, we calculated the ROC integral
(on strictly test data) as a function of the number of genes considered. We used a greater number of
iterations for this analysis because training with a small number of points tends to be noisy. Thus, this
approach allowed us to examine which factor contributes more to the performance of the algorithm:
measuring more cells or more properties.

\subsection*{\normalsize Chemotherapeutic response}

In the second part of this study, we ask whether actual drugs act as classifiers. More specifically, we
predict drug sensitivity of cell lines using gene expression data. To do this, we use the same GLM
framework but now with a normal distribution because the original dataset defines sensitivity (
$-log(GI_{50}))$ as a continuous variable. As above, the GLM parameter for each gene reflects how well it
predicts drug sensitivity. Modeling drug response as a GLM allowed us to ask whether drugs act like
classifiers by using gene expression to determine their effects on cells.

$$
\text{drug response}(\bm{\beta},\mathbf{x}) = \mathbf{x}\bm{\beta} = \beta_0 + \sum_k x_k \beta_k
$$

To test how well the predictions generalized to naïve data, we again used two levels of CV as in section I.
We measured the performance of the algorithm with an \emph{R value} (rather than an ROC curve) because
drug sensitivity is a continuous variable. As above, we quantified the performance of the algorithm as a
function of the number of genes considered. This allowed us to measure to what degree actual drugs act
as classifiers.

To more clearly visualize drug sensitivity's dependence on gene expression, we produced drug sensitivity
heatmaps. We ranked genes based upon the parameters returned by the GLM, chose the best two, and
then plotted drug sensitivity (``heat'') as a function of expression of each of the two genes. To smooth the
heatmap, we used the Matlab function griddata.

\subsection*{\normalsize Treatment optimization}

In the third part of this study we propose a method to optimize cancer treatment using the GLM
framework. This consists of choosing the next best molecular target for a specified cancer drug. To do
this, we build a classifier that discriminates between malignant and benign cancer (or ideally, healthy and
cancer). We use only two features in this classifier: the drug's behavior, $x_{drug}$
and that of a single other gene, $x_{marker}$. Thus, in terms of the marker levels, $\mathbf{x}$, and their weights,
$\bm{\alpha}$ which are distinct from $\bm{\beta}$ above):

$$
\text{cancer class}(\bm{\alpha},\mathbf{x}) = \mathbf{x} \bm{\alpha} = \alpha_0 + x_{drug}\alpha_{drug} + x_{marker}\alpha_{marker}
$$

Where $x_{drug}$ is given by how the drug's behavior depends on gene expression (a ``composite gene''),
found as in section II.

$$
x_{drug} = \mathbf{x}\bm{\beta} = \beta_0 + \sum_k x_k \beta_k
$$

We then iterated through many possible markers to determine which one allowed the best discrimination
(highest AUC) between invasive and non-invasive cancer when combined with the drug. Rather than
iterate through all possible markers (nearly 20,000), we identified 200 markers that best correlated with
invasiveness and iterated only through those. We then trained the classifier and tested it using two levels
of CV: 10-fold CV to determine the regularization penalty parameter and 3-fold CV for testing. We
defined the best marker as the one that led to the classifier with the highest AUC with ROC analysis.

\section*{\large Acknowledgments}

We thank Joe Gray for providing the experimental data used in sections II and III.

\section*{\large Funding sources}

PL was funded by NIH Grant 5P01NS044393. KK was funded by NIH Grant R01NS063399. RR and MR were funded by the University of Chicago Women's Board. TK was funded by the Machiah Foundation.

\bibliographystyle{plos2009}
\bibliography{cancerpaper}

\begin{thebibliography}{10}
\providecommand{\url}[1]{\texttt{#1}}
\providecommand{\urlprefix}{URL }
\expandafter\ifx\csname urlstyle\endcsname\relax
  \providecommand{\doi}[1]{doi:\discretionary{}{}{}#1}\else
  \providecommand{\doi}{doi:\discretionary{}{}{}\begingroup
  \urlstyle{rm}\Url}\fi
\providecommand{\bibAnnoteFile}[1]{%
  \IfFileExists{#1}{\begin{quotation}\noindent\textsc{Key:} #1\\
  \textsc{Annotation:}\ \input{#1}\end{quotation}}{}}
\providecommand{\bibAnnote}[2]{%
  \begin{quotation}\noindent\textsc{Key:} #1\\
  \textsc{Annotation:}\ #2\end{quotation}}
\providecommand{\eprint}[2][]{\url{#2}}

\bibitem{cite1}
Landau Da, Carter SL, Stojanov P, McKenna A, Stevenson K, et~al. (2013)
  {Evolution and impact of subclonal mutations in chronic lymphocytic
  leukemia.}
\newblock Cell 152: 714--26.
\newblock \doi{10.1016/j.cell.2013.01.019}.
\bibAnnoteFile{cite1}

\bibitem{cite2}
Bishop CM, Others (2006) {Pattern recognition and machine learning}, volume~4.
\newblock springer New York.
\bibAnnoteFile{cite2}

\bibitem{cite3}
Hanahan D, Weinberg Ra (2011) {Hallmarks of cancer: the next generation.}
\newblock Cell 144: 646--74.
\newblock \doi{10.1016/j.cell.2011.02.013}.
\bibAnnoteFile{cite3}

\bibitem{cite4}
{van 't Veer} LJ, Dai H, van~de Vijver MJ, He YD, Hart AAM, et~al. (2002) {Gene
  expression profiling predicts clinical outcome of breast cancer.}
\newblock Nature 415: 530--6.
\newblock \doi{10.1038/415530a}.
\bibAnnoteFile{cite4}

\bibitem{cite5}
Yun J, Frankenberger Ca, Kuo WL, Boelens MC, Eves EM, et~al. (2011) {Signalling
  pathway for RKIP and Let-7 regulates and predicts metastatic breast cancer.}
\newblock The EMBO journal 30: 4500--14.
\newblock \doi{10.1038/emboj.2011.312}.
\bibAnnoteFile{cite5}

\bibitem{cite6}
Ramaswamy S, Tamayo P, Rifkin R, Mukherjee S, Yeang CH, et~al. (2001)
  {Multiclass cancer diagnosis using tumor gene expression signatures.}
\newblock Proceedings of the National Academy of Sciences of the United States
  of America 98: 15149--54.
\newblock \doi{10.1073/pnas.211566398}.
\bibAnnoteFile{cite6}

\bibitem{cite7}
Sch{\"o}lkopf B, Smola AJ (2002) Learning with kernels.
\newblock “The” MIT Press.
\bibAnnoteFile{cite7}

\bibitem{cite8}
Pearson K (1901) {On lines and planes of closest fit to systems of points in
  space}.
\newblock The London, Edinburgh, and Dublin Philosophical \ldots : 559--572.
\bibAnnoteFile{cite8}

\bibitem{cite9}
Quackenbush J (2001) {Computational analysis of microarray data.}
\newblock Nature reviews Genetics 2: 418--27.
\newblock \doi{10.1038/35076576}.
\bibAnnoteFile{cite9}

\bibitem{cite10}
Khan J, Wei JS, Ringn\'{e}r M, Saal LH, Ladanyi M, et~al. (2001)
  {Classification and diagnostic prediction of cancers using gene expression
  profiling and artificial neural networks.}
\newblock Nature medicine 7: 673--9.
\newblock \doi{10.1038/89044}.
\bibAnnoteFile{cite10}

\bibitem{cite11}
Staunton JE, Slonim DK, Coller Ha, Tamayo P, Angelo MJ, et~al. (2001)
  {Chemosensitivity prediction by transcriptional profiling.}
\newblock Proceedings of the National Academy of Sciences of the United States
  of America 98: 10787--92.
\newblock \doi{10.1073/pnas.191368598}.
\bibAnnoteFile{cite11}

\bibitem{cite12}
Potti A, Dressman HK, Bild A, Riedel RF, Chan G, et~al. (2006) {Genomic
  signatures to guide the use of chemotherapeutics.}
\newblock Nature medicine 12: 1294--300.
\newblock \doi{10.1038/nm1491}.
\bibAnnoteFile{cite12}

\bibitem{cite13}
Sotiriou C, Piccart MJ (2007) {Taking gene-expression profiling to the clinic:
  when will molecular signatures become relevant to patient care?}
\newblock Nature reviews Cancer 7: 545--53.
\newblock \doi{10.1038/nrc2173}.
\bibAnnoteFile{cite13}

\bibitem{cite14}
van't Veer LJ, Bernards R (2008) {Enabling personalized cancer medicine through
  analysis of gene-expression patterns.}
\newblock Nature 452: 564--70.
\newblock \doi{10.1038/nature06915}.
\bibAnnoteFile{cite14}

\bibitem{cite15}
Dalerba P, Kalisky T, Sahoo D, Rajendran PS, Rothenberg ME, et~al. (2011)
  {Single-cell dissection of transcriptional heterogeneity in human colon
  tumors.}
\newblock Nature biotechnology 29: 1120--7.
\newblock \doi{10.1038/nbt.2038}.
\bibAnnoteFile{cite15}

\bibitem{cite16}
Tibshirani R (1996) Regression shrinkage and selection via the lasso.
\newblock Journal of the Royal Statistical Society Series B (Methodological) :
  267--288.
\bibAnnoteFile{cite16}

\bibitem{cite17}
Heiser LM, Sadanandam A, Kuo WL, Benz SC, Goldstein TC, et~al. (2012) {Subtype
  and pathway specific responses to anticancer compounds in breast cancer.}
\newblock Proceedings of the National Academy of Sciences of the United States
  of America 109: 2724--9.
\newblock \doi{10.1073/pnas.1018854108}.
\bibAnnoteFile{cite17}

\bibitem{cite18}
Freund Y, Schapire R (1995) {A desicion-theoretic generalization of on-line
  learning and an application to boosting}.
\newblock Computational learning theory \doi{10.1007/3-540-59119-2\_166}.
\bibAnnoteFile{cite18}

\bibitem{cite19}
Yachida S, Jones S, Bozic I, Antal T, Leary R, et~al. (2010) {Distant
  metastasis occurs late during the genetic evolution of pancreatic cancer.}
\newblock Nature 467: 1114--7.
\newblock \doi{10.1038/nature09515}.
\bibAnnoteFile{cite19}

\bibitem{cite20}
Gerlinger M, Rowan AJ, Horswell S, Larkin J, Endesfelder D, et~al. (2012)
  {Intratumor heterogeneity and branched evolution revealed by multiregion
  sequencing.}
\newblock The New England journal of medicine 366: 883--92.
\newblock \doi{10.1056/NEJMoa1113205}.
\bibAnnoteFile{cite20}

\bibitem{cite21}
Lee MJ, Ye AS, Gardino AK, Heijink AM, Sorger PK, et~al. (2012) {Sequential
  application of anticancer drugs enhances cell death by rewiring apoptotic
  signaling networks.}
\newblock Cell 149: 780--94.
\newblock \doi{10.1016/j.cell.2012.03.031}.
\bibAnnoteFile{cite21}

\bibitem{cite22}
Tan A, Gilbert D (2003) {Ensemble machine learning on gene expression data for
  cancer classification} 2: 1--10.
\bibAnnoteFile{cite22}

\bibitem{cite23}
Bozic I, Reiter JG, Allen B, Antal T, Chatterjee K, et~al. (2013) {Evolutionary
  dynamics of cancer in response to targeted combination therapy}.
\newblock eLife 2: e00747--e00747.
\newblock \doi{10.7554/eLife.00747}.
\bibAnnoteFile{cite23}

\bibitem{cite24}
Fu X, Rivera A, Tao L, {De Geest} B, Zhang X (2012) {Construction of an
  oncolytic herpes simplex virus that precisely targets hepatocellular
  carcinoma cells.}
\newblock Molecular therapy : the journal of the American Society of Gene
  Therapy 20: 339--46.
\newblock \doi{10.1038/mt.2011.265}.
\bibAnnoteFile{cite24}

\end{thebibliography}

\end{document}